\documentclass[twocolumn,aps,prb,showpacs]{revtex4}

\begin{document}
\author{Yaroslav Tserkovnyak and Arne Brataas} \affiliation{Harvard
University, Lyman Laboratory of Physics, Cambridge, Massachusetts 02138}

\title{Current and spin torque in double tunnel barrier ferromagnet--superconductor--ferromagnet systems}

\begin{abstract}
We calculate the current and the spin torque in small symmetric double
tunnel barrier ferromagnet--superconductor--ferromagnet (\textit{F-S-F})
systems. Spin accumulation on the
superconductor governs the transport properties when the spin-flip
relaxation time is longer than the transport dwell time. In the
elastic transport regime, it is demonstrated that the relative change
in the current (spin torque) for \textit{F-S-F} systems equals the relative
change in the current (spin torque) for
ferromagnet--normal metal--ferromagnet (\textit{F-N-F}) systems upon changing
the relative magnetization direction of the two ferromagnets. This
differs from the results in the inelastic transport regime where
spin accumulation suppresses the superconducting gap and dramatically
changes the magnetoresistance [S.\ Takahashi, H.\ Imamura, and S.\
Maekawa, Phys.\ Rev.\ Lett.\ {\bf 82}, 3911 (1999)]. The experimental
relevance of the elastic and inelastic transport regimes,
respectively, as well as the reasons for the change in the transport
properties are discussed.
\end{abstract}

\date{\today}


\maketitle

\section{Introduction}
\label{Introduction}

Conventional low-temperature superconductors result from the pairing
of electrons with spin up and spin down and opposite momentum.
Ferromagnets in contact with a nonmagnetic metal can induce a
nonequilibrium spin accumulation in the nonmagnetic metal. Such a
spin accumulation in a superconductor can reduce the accessible number
of pairs of spin-up and spin-down electrons. Consequently,
spin accumulation in superconductors can reduce the superconducting
gap and dramatically change the transport properties.

Spin accumulation and its effect on the current-voltage
characteristics\cite{Barnas:epl98,Brataas:epjb99}
and the shot noise\cite{Tserkovnyak:prb01} have been studied thoroughly
in double barrier \textit{F-N-F} systems.
We will here consider double tunnel barrier \textit{F-S-F} systems
with general noncollinear magnetization directions.
The influence of spin accumulation on
the superconducting gap, the current and the spin torque will be
considered.

Spin accumulation and its influence on the superconducting gap
strongly depend on the competition between different relevant
transport time scales; the transport dwell time $\tau_{d}$
characterizes the typical time an electron spends on the
superconducting island on passing through the system, the spin-flip
relaxation time $\tau _{\text{sf}}$ gives the time-scale for the
coherence of the electron spin, and the energy relaxation time
$\tau_{E}$ denotes the time scale for the interchange of energy
between the quasiparticle and the rest of the system.
Spin accumulation in \textit{F-N-F} or \textit{F-S-F} systems requires that the
spin-flip relaxation time is much longer than the transport dwell
time, $\tau_{\text{sf}}\gg\tau_{d}$. The transport dwell time $\tau_d$
decreases with decreasing size of the system via the density of states
and consequently spin accumulation will only take place in
mesoscopically small systems (for a discussion, see
Ref.~\onlinecite{Brataas:epjb99}).  \textit{F-S-F} systems have been studied in
an interesting article\cite{Takahashi:prl99}
in the limit when $\tau_{\text{sf}}\gg \tau_{d}\gg \tau_{E}$,
\textit{i.e.}, when the energy relaxation
is sufficiently strong so that the quasiparticles for each spin
relax to Fermi-Dirac (FD) distribution functions with spin-dependent
chemical potentials. We will below show that a different regime
$(\tau_E,\tau_{\text{sf}}) \gg \tau_d$ is relevant in some
experimental situations, and we will work out the transport properties
and compare them to the inelastic regime.

Our aim is to provide a complete description of the elastic transport
regime for \textit{F-S-F} systems with noncollinear magnetization directions.
We will also generalize the results in the inelastic transport
regime\cite{Takahashi:prl99} to noncollinear magnetization
configurations. Both the current and the spin torque relevant for
magnetic switching behavior in small ferromagnetic
particles\cite{Sloncz:mmm95,Myers:sc00} will be studied.

\section{Transport regimes}
\label{Regime}

The relevance of the elastic or inelastic transport regimes depends on
two time scales intrinsic to the superconducting island, the energy
relaxation time $\tau_E$ and the spin-flip relaxation time
$\tau_{\text{sf}}$ as well as the time-scale governed by the contacts
to the superconducting island, $\tau_d$. As discussed in the
Introduction, spin accumulation requires $\tau_{\text{sf}}\gg\tau_d$
and we will assume that this is satisfied in the following.
This implies that when the intrinsic time-scales are in the regime
$\tau_E\gg\tau_{\text{sf}}$, the spin accumulation requirement dictates
the elastic transport regime, $\tau_E\gg\tau_d$.
In the opposite limit, $\tau_E\ll\tau_{\text{sf}}$ both elastic and
inelastic regimes can be relevant depending on the contacts.
Let us now discuss the ratio between the spin-flip relaxation time
$\tau_{\text{sf}}$ and the energy relaxation time $\tau_E$ for
conventional s-wave superconductors. First, let us consider the case of Al.

The energy relaxation rate $1/\tau_E$ has contributions due to the
electron-electron scattering ($1/\tau_{e-e}$) and due to the electron-phonon
processes, \textit{i.e.}, quasiparticle-phonon scattering
and quasiparticle recombination with phonon emission to form Cooper pairs,
($1/\tau_{e-\text{ph}}$).
It is usually assumed that the two contributions add up:
$1/\tau_E=1/\tau_{e-e}+1/\tau_{e-\text{ph}}$.
In most conventional superconductors, the dominant quasiparticle energy
relaxation rate is due to the electron-phonon processes, except for
metals with a large Debye frequency and a small superconducting gap,
such as Al and Zn, for which electron-electron scattering processes can be
significant.\cite{Kaplan:prb76}

The energy relaxation time due to the electron-phonon processes was studied in
Ref.~\onlinecite{Kaplan:prb76} for a variety of materials. For Al,
$\tau_{e-\text{ph}}\approx5\times10^{-8}$~s close to the Fermi level
at the critical temperature. A further reduction of the temperature
leads to an even smaller rate for the electron-phonon processes at the
relevant energy $\Delta(T)$.\cite{Kaplan:prb76} Furthermore, the electron recombination rate to form Cooper pairs is reduced for spin-polarized quasiparticles.

The energy relaxation time due to the electron-electron interaction
in dirty normal state systems at an energy $\epsilon$ with respect to
the Fermi level is\cite{Altshuler:jetp79}
\begin{equation}
\tau_{e-e}=\frac{8\pi (k_{F}\lambda)^{2}}{\sqrt{6}(\epsilon /\hbar
)^{3/2}\tau ^{1/2}} \, ,
\label{ee}
\end{equation}
where $k_{F}$ is the Fermi wave vector,
$\tau$ is the elastic scattering time, and $\lambda=v_{F}\tau $ is
the mean free path. Relevant energy scales for \textit{F-S-F} systems are
around the superconducting gap (in Al, $ T_{c}\approx 1.2$~K) and lower.
Using a typical Fermi energy for metallic systems $E_F=10$~eV
and a value of the electron mean free path for a dirty system
$k_F\lambda=10$ (which corresponds to $\lambda\approx6$~\AA)
we find for Al $\tau_{e-e}\approx4\times10^{-7}$~s.
For cleaner systems, $\tau_{e-e}$ can be much larger since
$\tau_{e-e}\propto\tau^{3/2}$.

The spin-flip relaxation time depends on the relativistic spin-orbit
interaction and the magnetic impurity scattering. For nonmagnetic
impurities, the spin-orbit interaction dominates the spin-flip rate.
The spin-orbit scattering time in
superconducting Al was found to be $\tau_{\text{so}}\approx2\times10^{-11}$~s in
the pioneering tunneling spectroscopy measurements by Tedrow and
Meservey.\cite{Meservey:pr94} In their experiment they used an Al film with the elastic scattering time $\tau\approx10^{-14}$~s
(corresponding to $\lambda\approx20$~nm) from which we can estimate
$\tau_{e-e}$ using Eq.~(\ref{ee}). For such a clean system,
the electron-electron interaction is weak and
$\tau_E\approx\tau_{e-\text{ph}}$. Johnson and Silsbee\cite{Johnson:prl88}
subsequently found a longer spin-flip relaxation time in an Al (normal
state) thin film, $\tau _{\text{sf}}\sim 10^{-8}$~s. To the best of our
knowledge there are no other experiments that have been so successful
in achieving such a long spin-flip relaxation time in Al.
Besides, the spin-flip
relaxation time decreases in small metal particles due to the enhanced
scattering rate at the surface\cite{Brataas:epjb99} and in materials
with magnetic impurities due to the spin-spin
scattering.\cite{Yang:prl94} We, therefore, consider the
long spin-flip relaxation $\tau_{\text{sf}}\sim 10^{-8}$~s an
optimistic estimate in small superconducting Al particles and believe
that times of the order $\tau_{\text{sf}}\sim 10^{-11}$~s are currently
easier to achieve experimentally.

The energy relaxation time can therefore be of the same order as the
spin-flip relaxation time or larger for superconducting
Al particles. It is therefore likely that at least
some experiments on \textit{F-S-F} systems with Al island will be in the regime
$\tau_{E}\gg\tau_{\text{sf}}\gg\tau_{d}$.
This motivates our study of the elastic transport regime.

Let us extend this discussion to the case of some other common
s-wave superconductors. Kaplan {\it et al.}\cite{Kaplan:prb76}
showed that for many simple
conventional superconductors (Al, Zn, Nb, In, Sn, Ta, Hg, Tl, Pb)
$\tau_{e-\text{ph}}$ ranges between $10^{-7}$~s and $10^{-11}$~s.
We have to compare it to the spin-orbit scattering time $\tau_{\text{so}}$.
Approximately, $\tau_{\text{so}}$ scales linearly with the elastic scattering
time $\tau$ according to\cite{Abrikosov:zetf62}
\begin{equation}
\tau/\tau_{\text{so}}\sim (\alpha Z)^4 \, ,
\label{ag}
\end{equation}
where $\alpha$ is the fine structure constant and $Z$ is the atomic number of the impurity atoms. Meservey and Tedrow\cite{Meservey:prl78} have showed by compilation of data that Eq.~(\ref{ag}) is a reasonable estimate of the spin-orbit scattering rate for very clean small particles or films, when the surface collisions dominate the elastic scattering processes. Nevertheless, they showed that for many metals with atomic numbers under 100, Eq.~(\ref{ag}) typically gives an underestimate of the scattering rate $1/\tau_{\text{so}}$ by a factor of the order of 10. According to Eq.~(\ref{ag}), in metals heavier than Al, it is even more difficult to achieve a lower spin-flip relaxation rate.

We therefore conclude that the elastic regime $(\tau_{E},\tau_{\text{sf}})\gg\tau_{d}$ can be achieved in experiments on \textit{F-S-F} systems with a variety of materials used for the superconducting node. This is true for relatively clean systems, when $\tau_{e-\text{ph}}\ll\tau_{e-e}$. For dirty systems, electron-electron interactions can become significant and since $\tau_{e-e}\propto\tau^{3/2}$ (\ref{ee}) and $\tau_{\text{so}}\propto\tau$ (\ref{ag}), $\tau_{e-e}/\tau_{\text{so}}\propto\sqrt{\tau}\rightarrow 0$ in the theoretical limit $\tau\rightarrow 0$. If this limit is physically achieved, both elastic and inelastic regimes can be relevant. The above discussion does not apply to high-$T_c$ superconductors which typically have much higher energy relaxation rate. In particular, some recent experiments\cite{Vasko:prl97} on spin-polarized quasiparticle injection in the cuprate superconductors indicate the relevance of the inelastic regime as studied in Refs.~\onlinecite{Takahashi:prl99} and \onlinecite{Yamashita:cm01}.

In the elastic regime, the ratio between the energy-relaxation time and the spin-flip relaxation time does not matter; the occupation of the quasiparticles states are not determined by FD distributions, but has to be determined from the transport equations. These transport equations will be found in the following.

\section{Model}
\label{Model}

The \textit{F-S-F} system consists of a superconducting island connected to a
left and a right ferromagnet by symmetric tunnel junctions. The left and the
right ferromagnets are attached to a left and a right reservoir by
good metallic contacts; it is assumed that the current is only limited
by the tunnel conductance between the ferromagnets and the
superconductor. Each monodomain ferromagnet has a well-defined
magnetization direction, ${\bf m}_l$ and ${\bf m}_r$ for the left and
the right ferromagnet, respectively. The system is biased by an
external voltage source between the left and the right reservoirs.
The convention is such that the chemical potential in the left reservoir
is higher than that in the
right reservoir by $eV$ ($e>0$). We will consider the
current and spin current from the superconducting island to the left reservoir,
and the spin torque on the left ferromagnet.

We consider the regime when the tunnel conductance is much larger
than the quantum conductance so that the Coulomb charging effects can be
disregarded. It is further assumed that the level spacing is much
smaller than the temperature and the superconducting gap, so that the
superconducting state can be well described by the BCS theory. The
dwell time is much larger than the time for the diffusion through the
superconducting island so that the phase-space occupation of the
quasiparticles can be described semiclassically. It is also assumed that the
spin-flip relaxation time is longer than the transport dwell time in
all calculations, $\tau_{\text{sf}} \gg \tau_{d}$, so that a
nonequilibrium spin accumulation on the superconducting island can
exist.

We describe the electron transport through a tunnel junction
sandwiched between a superconductor and a ferromagnet using a
phenomenological tunneling Hamiltonian. The total Hamiltonian of the
system is $H=H_0+H^\prime$, where
\begin{equation}
H_0=\sum_k a_k^\dagger E_k a_k+\sum_p c_p^\dagger \hat{\epsilon}_p c_p
\end{equation}
is the Hamiltonian of the uncoupled superconductor and ferromagnet and
\begin{equation}
H^\prime=\sum_{kp}c_p^\dagger\hat{T}_{kp}a_k+\text{H.c.}
\label{Ht}
\end{equation}
is the phenomenological tunneling Hamiltonian. The ferromagnet has
a general magnetization direction ${\bf m}$.
$c_p$ ($a_k$) is a column vector of
spin-up and spin-down annihilation operators for an electron with
a momentum $p$ ($k$) in the ferromagnet (superconductor),
$\hat{T}_{kp}=T_{kp\uparrow}\hat{u}^\uparrow+T_{kp
\downarrow}\hat{u}^\downarrow$ and
$\hat{\epsilon}_p=\epsilon_{p\uparrow}
\hat{u}^\uparrow+\epsilon_{p\downarrow}\hat{u}^\downarrow$.  The
tunneling matrix $\hat{T}$ and the ferromagnet energy matrix
$\hat{\epsilon}$ depend on the magnetization direction of the
ferromagnet through the projection matrices $\hat{u}^{\uparrow
}=\left(\hat{1}+\hat{\text{\boldmath$\sigma $}}\cdot {\bf m}\right) /2$ and
$\hat{u}^{\downarrow }=\left(\hat{1}-\hat{\text{\boldmath$ \sigma $}}\cdot
{\bf m}\right) /2$, where ${\bf m}$ is a unit vector in the direction
of the local magnetization and $\hat{\text{\boldmath$\sigma $}}=(
\hat{\sigma}_{x},\hat{\sigma}_{y},\hat{\sigma}_{z})$ is a vector of
the $ 2\times 2$ Pauli matrices.

The $ 2\times 2$ current matrix in the spin space evaluated inside the
superconductor is
\begin{equation}
\hat{I}^{\alpha\beta}=e\frac{i}{\hbar}\sum_{k}\left\langle\left[
H^\prime,a_{\beta k}^\dagger a_{\alpha k}\right]\right\rangle=e\frac{i}{
\hbar}\mbox{Tr}(\hat{M}\hat{T})+\text{H.c.} \, ,
\label{I}
\end{equation}
where $\hat{M}_{kp}^{\alpha\beta}=\langle c_{\beta p}^\dagger a_{\alpha
k}\rangle$ and the trace is taken over the momentum space indices.
The current is $I =\sum_{\alpha}\hat{I}^{\alpha\alpha}$
and the spin current is $\hat{\bf I}_{s}=-\hbar/(2e)
\sum_{\alpha\beta}\hat{\text{\boldmath$\sigma $}}_{\alpha\beta }
\hat{I}^{\beta \alpha }$.

In order to calculate $\hat{M}$, we express the electron operators $a$ in
terms of the quasiparticle operators $\gamma$ by the Bogoliubov transformation:
\begin{equation}
a_k=\hat{u}_k\gamma_k+\hat{v}_k\gamma_{-k}^* \, ,
\label{Bt}
\end{equation}
where 
\[
\hat{u}_k=u_k\left( 
\begin{array}{cc}
1 & 0 \\ 
0 & 1
\end{array}
\right),~ \hat{v}_k=v_k\left( 
\begin{array}{cc}
0 & 1 \\ 
-1 & 0
\end{array}
\right) 
\]
and $u_k=\left[(1+\xi_k/E_k)/2\right]^{1/2}$, $v_k=\left[(1-\xi_k/E_k)/2\right]^{1/2}$, where $\xi_k$ is the electron energy relative to the
chemical potential, $E_k=\sqrt{\xi_k^2+\Delta^2}$ is the corresponding quasiparticle energy and $\Delta$ is the superconducting gap.

Up to this point, everything is exact within the tunneling Hamiltonian
phenomenology. We calculate $\hat{M}$ [and thus find the current
(\ref{I})] to the lowest nonvanishing order in $\hat{T}$. This approximation
neglects Andreev reflection and higher-order corrections to the
quasiparticle current. In the next section, we show the final result
for the current (\ref{I}) and apply it to describe the transport
properties of double tunnel barrier \textit{F-S-F} systems.

\section{Theoretical results}
\label{Results}

The current operator is a $2\times 2$ matrix in the spin space. Let
us first include the result when the island is in the normal state.
It was recently demonstrated\cite {Brataas:prl00} that
the $2\times 2$ current (per unit of energy, at a given energy $\epsilon $)
through a {\em single} normal metal--ferromagnet (\textit{N-F}) junction
at the normal metal side can be written
as
\begin{eqnarray}
\hat{\imath}_N\left( \epsilon \right)e &=&g^{\uparrow }\hat{u}^{\uparrow
}\left. \delta \hat{f}_N\left( \epsilon \right) \right.
\hat{u}^{\uparrow }+g^{\downarrow }\hat{u}^{\downarrow }\left. \delta
\hat{f}_N\left( \epsilon \right) \right. \hat{u}^{\downarrow }
\nonumber \\ &&+g^{\uparrow \downarrow }\hat{u}^{\uparrow }\left.
\delta \hat{f} _N\left( \epsilon \right) \right. \hat{u}^{\downarrow
}+g^{\downarrow \uparrow }\hat{u}^{\downarrow }\left. \delta
\hat{f}_N\left( \epsilon \right) \right. \hat{u}^{\uparrow } \, ,
\label{iN}
\end{eqnarray}
where $\delta \hat{f}_N\left( \epsilon \right) $ is the $2\times 2$
matrix difference
in the distribution function on the ferromagnetic and the normal metal
sides $ \delta \hat{f}_N\left( \epsilon \right) =\hat{f}_F\left(
\epsilon \right) -\hat{f}_N\left( \epsilon \right) $. The
semiclassical distribution function $\hat{f}\left( \epsilon \right) $
is defined by
\begin{equation}
\left\langle a_{\alpha m}^{\dagger }\left( \epsilon \right) a_{\beta
n}\left( \epsilon ^{\prime }\right) \right\rangle =\delta _{nm}\delta
\left( \epsilon -\epsilon ^{\prime }\right) \hat{f}^{\beta \alpha
}\left( \epsilon \right) \, ,
\end{equation}
where $a_{\alpha m}^{\dagger }$ creates a spin-$\alpha $ particle in
the $m^{ \text{th}}$ quantum state. In Eq.\ (\ref{iN}), $g^{\uparrow
}$ ($g^{\downarrow }$) denotes the junction conductance for a spin-up
(spin-down) electron, and $g^{\uparrow \downarrow }$ ($=\left(
g^{\downarrow \uparrow }\right) ^{\ast } $) is the mixing conductance
first introduced in Ref.~\onlinecite{Brataas:prl00}. For a tunnel
junction, as we study here, we find that the calculation based on the
tunneling Hamiltonian (\ref{Ht}) gives the mixing conductance $
g^{\uparrow \downarrow }=\left( g^{\uparrow }+g^{\downarrow }\right)
/2$, which is consistent with the alternative tunneling Hamiltonian
phenomenology in Ref.~\onlinecite{Brataas:prl00}.

The current through the \textit{F-N-F} system is found by using the charge and spin
conservation on the normal metal node, so that the current through the
first junction cancels the current through the second junction; in the
elastic transport regime, the charge and spin are conserved at each
energy: $\hat{\imath}_{1}(\epsilon )+\hat{\imath} _{2}(\epsilon )=0$,
and in the inelastic scattering regime, the distribution functions are
assumed to be FD distributions for each spin direction and the total
charge and spin are conserved:
$\int_{-\infty}^{\infty}\text{d}\epsilon
\left[\hat{\imath}_{1}(\epsilon)+\hat{\imath}_{2}(\epsilon)\right]=0$.
By using the current conservation, the distribution on the normal metal
node can be found, and, by inserting the resulting distribution into
Eq.~(\ref{iN}), the resulting current through the tunnel junction is
\cite{Brataas:prl00}
\begin{equation}
I(V)=\left[ 1-p^2\sin^2\theta/2 \right]\frac{g_{\uparrow }+g_{\downarrow }}{2}V \, .
\end{equation}
The current only depends on the total conductance of spin-up and
spin-down electrons $g_{\uparrow}+g_{\downarrow}$ and the relative
polarization of the contacts
$p=(g_{\uparrow}-g_{\downarrow})/(g_{\uparrow}+g_{\downarrow})$ and
has a simple $\sin^2\theta/2$ depends on the relative magnetization
angle $\theta$.\cite{Brataas:prl00} For double barrier
\textit{F-N-F} systems with general contacts a
more general and complicated angular dependence is
found.\cite{Brataas:prl00}

\begin{widetext}
In this paper we generalize Eq.~(\ref{iN}) to describe quasiparticle
current through conventional s-wave superconductor--ferromagnet (\textit{S-F})
junctions. We find that the quasiparticle current
$\hat{\imath}_{\text{qp}}$ can similarly be expressed by
Eq.~(\ref{iN}) by replacing $\delta\hat{f}_N $ with an effective
difference in the distribution functions $\delta\hat{f}_S$ and renormalizing
the conductance by the BCS density of states:
\begin{eqnarray}
\hat{\imath}_{\text{qp}}(\epsilon)e=N_S(\epsilon)\left[g^\uparrow\hat{u}^\uparrow\delta\hat{f}_S(\epsilon)\hat{u}^\uparrow+g^\downarrow\hat{u}^\downarrow\delta\hat{f}_S(\epsilon)\hat{u}^\downarrow+g^{\uparrow\downarrow}\hat{u}^\uparrow\delta\hat{f}_S(\epsilon)\hat{u}^\downarrow+g^{\downarrow\uparrow}\hat{u}^\downarrow\delta\hat{f}_S(\epsilon)\hat{u}^\uparrow\right] \, ,
\label{iS}
\end{eqnarray}
where
\begin{equation}
\delta \hat{f}_S\left( \epsilon \right)  = 
\left\{
\begin{array}{cl}
\hat{f}_F\left( \epsilon \right) -\frac{1}{2}\left(\hat{f}^>_S\left( \epsilon
\right) +\hat{f}^<_S\left( \epsilon \right) \right) -\frac{Q(\epsilon)}{2}\left(
\hat{f}^>_S\left( \epsilon \right) -\hat{f}^<_S\left( \epsilon \right)
\right)
& ,~\epsilon >\Delta \nonumber \\
\hat{f}_F\left( \epsilon \right) -1+\frac{1}{2}\left(\hat{\mathcal{F}}
^>_S\left( -\epsilon \right) +\hat{\mathcal{F}}^<_S\left( -\epsilon
\right) \right) + \frac{Q(\epsilon)}{2}\left( \hat{\mathcal{F}}^>_S\left( -\epsilon \right)
-\hat{\mathcal{F}}^<_S\left( -\epsilon \right) \right)
& ,~\epsilon <-\Delta 
\end{array}
\right.
\label{fS}
\end{equation}
Here, $N_S(\epsilon)=|\epsilon|/\sqrt{\epsilon^2-\Delta^2}$
is the normalized BCS density of states,
$Q(\epsilon)=\sqrt{\epsilon
^{2}-\Delta ^{2}}/\epsilon $ is the quasiparticle effective charge,
$\hat{f}^>_S$ ($ \hat{f}^<_S$) is the superconducting distribution
function of electron-like (hole-like) quasiparticles (in the
superconductor, the distribution function is defined via quasiparticle
operators $\gamma _{\alpha m}$ related to electron operators
$a_{\alpha m}$ by the Bogoliubov transformation
(\ref{Bt})). $\hat{\mathcal{F}}$ is a ``flipped'' distribution
function $\hat{f}$; that is if $
\hat{f}=f_{0}\hat{1}+f_{s}\hat{\text{\boldmath$\sigma $}}\cdot {\bf u}$, then
$\hat{\mathcal{F}}=f_{0}\hat{1}-f_{s}\hat{\text{\boldmath$\sigma $}}\cdot
{\bf u}$. The energy $\epsilon$ is measured with respect to
the chemical potential of the Cooper pair condensate.
\end{widetext}

It is important to note that $\epsilon$ in Eqs.~(\ref{iS}) and (\ref{fS})
is the electron energy and
thus can be positive or negative; the quasiparticle energy in the
superconductor (which we will denote $E$), on the other hand,
is always positive.

The superconducting gap is determined by the (nonequilibrium)
occupation of the quasiparticles:
\begin{equation}
\ln\left(\frac{\Delta_0}{\Delta}\right)=\int_{\Delta }^{\infty}
\text{d}E\frac{\mbox{Tr}\left[\hat{f}_S(E)\right]}{\sqrt{E^{2}-\Delta^{2}}} \, ,
\label{sc}
\end{equation}
where $\hat{f}_S(E)=\left[\hat{f}^>_S(E)+\hat{f}^<_S(E)\right]/2$ is the
distribution function for combined electron-like and hole-like
excitations and $\Delta_0$ is the zero-temperature equilibrium gap.

In the elastic case ($\tau_{E} \gg\tau_{d}$), we solve for the
distribution function in the superconductor based on the conservation
of charge and spin on the superconducting node at {\em every} energy
$\epsilon$.  The resulting distribution function depends on the
spin-up and spin-down junction conductances, voltage bias,
superconducting gap $\Delta_{\text{el}}$ and the relative
magnetization direction of the
ferromagnets.  On the other hand, we find that the average occupation
number of a particular quasiparticle energy level $E>\Delta_{\text{el}}$ on the
superconducting island is simply 
\begin{equation}
\mbox{Tr}[\hat{f}_{S,\text{el}}(E)]=f(E-eV/2)+f(E+eV/2) \, ,
\label{Trf}
\end{equation}
where $f(E)$ is the Fermi function.
Eq.~(\ref{Trf}) is a direct consequence of the charge conservation at electron
energies $\epsilon=\pm E$.

The gap $\Delta_{\text{el}}$ is found self-consistently by inserting Eq.~(\ref{Trf})
into Eq.~(\ref{sc}):
\begin{eqnarray}
\ln\left(\frac{\Delta_0}{\Delta_{\text{el}}}\right)&=&\int_{\Delta_{\text{el}} }^{\infty}
\frac{\text{d}E}{\sqrt{E^{2}-\Delta_{\text{el}}^{2}}}  \nonumber \\
&&\times\left[f(E-eV/2)+f(E+eV/2)\right] \, .
\label{sg}
\end{eqnarray}
The gap is thus independent on the magnetization directions and the
polarization of the junctions. This is very different from the results
in the inelastic transport regime studied in Ref.~\onlinecite{Takahashi:prl99}.
The gap equation (\ref{sg}) gives the
same gap suppression as for normal metal--superconductor--normal
metal (\textit{N-S-N}) systems, {\it e.g.} as for a \textit{F-S-F} system with no
spin polarization. The properties of \textit{N-S-N} systems are well
known:\cite{Heslinga:prb93} At $T=0$, for $|eV/2|<\Delta_0$, electrons cannot
tunnel onto the superconducting island and the gap is invariant, but
for $|eV/2|>\Delta_0$, electrons tunnel and induce an excess number of
quasiparticles on the island which destroy the superconductivity
altogether.  Consequently, at $T=0$,
$\Delta_{\text{el}}(V)=\Delta_0\Theta(\Delta_0-|eV/2|)$, where $\Theta$ is the
Heaviside step-function.  Increasing the temperature decreases
$\Delta_{\text{el}}(V)$ and the critical voltage for the
superconductor--normal metal phase transition.
Moreover, $\Delta_{\text{el}}(V)$ remains
single-valued. This is in contrast to the inelastic transport regime
when the gap $\Delta_{\text{in}}(V)$ is multiple-valued for some
values of $V$ and $T$ and
the system exhibits hysteresis, so that the thermodynamic properties
of the island depend not only on the applied voltage but also on the
history of the system.

In the elastic transport regime, the magnetization direction dependent
quasiparticle current is
\begin{equation}
I_{\text{el}}(\theta,V,T)=\left[1-p^{2}\sin^2\theta/2\right]I_0\left(V,T,\Delta_{\text{el}}(V,T)\right) \, ,
\label{Iel}
\end{equation}
where $I_0(V,T,\Delta)$ is the current in the parallel configuration ($\theta=0$) of the magnetization directions and the gap $\Delta_{\text{el}}$ has to be determined self-consistently at each voltage bias according to Eq.~(\ref{sg}). In the parallel magnetization alignment, there is no spin accumulation and the current $I_0$ is given by the familiar result for \textit{N-S-N} systems with the total \textit{N-S} junction conductance $g^\uparrow+g^\downarrow$:
\begin{eqnarray}
I_0(V,T,\Delta)&=&(g^\uparrow+g^\downarrow)/e
\int_{\Delta}^{\infty}\text{d}EN_S(E) \nonumber \\
&&\times\left[f(E-eV/2)-f(E+eV/2)\right] \, . \nonumber \\
\label{I0}
\end{eqnarray}

Let us compare the results above in the elastic transport regime to
the transport properties in the inelastic regime
($\tau_{E}\ll\tau_d$). The inelastic transport
regime has been studied when the magnetization directions are
collinear.\cite{Takahashi:prl99} It was shown that in the
parallel (P) alignment of the magnetizations,
there is no bias dependence of the gap
(that is the gap $\Delta_{\text{in}}$ always equals
its equilibrium value $\Delta(T)$)
and the quasiparticles are distributed according to
the equilibrium FD distribution. In the antiparallel (A) alignment,
on the other hand, the gap is suppressed.\cite{Takahashi:prl99} This
is because the total (integrated over all energies) spin and charge
conservation on the superconducting island requires the chemical
potentials of the spin-up and spin-down quasiparticles to be shifted
oppositely by a finite value $\delta\mu$ with respect to the chemical
potential of the condensate. $\delta\mu/\Delta_0$ is a function of $
V/\Delta_0$, $T/\Delta_0$ and the ferromagnet polarization
$p$. Knowledge of this function gives the full thermodynamic and
transport description of the \textit{F-S-F} system in this configuration.
\cite{Takahashi:prl99} In particular,
\begin{equation}
\mbox{Tr}[\hat{f}_{S,\text{in}}(E)]=f(E-\delta\mu)+f(E+\delta\mu) \, ,
\label{Trfi}
\end{equation}
in contrast to Eq.~(\ref{Trf}),
and the gap $\Delta_{\text{in}}(V,T,p)$ can be found from the
self-consistency equation (\ref{sc}).

Let us generalize the result in Ref.~\onlinecite{Takahashi:prl99}
to noncollinear magnetization configurations.
For an intermediate alignment of the ferromagnet magnetizations, we
can find the magnitude of spin accumulation $\delta\mu$ and a
unit vector ${\bf u}$ (the direction of the nonequilibrium magnetization)
which diagonalizes the quasiparticle distribution function
on the island in the spin space, by requiring conservation of the total
spin and charge on the superconducting node. We find that ${\bf u}$
points along the relative magnetization of the two ferromagnets,
${\bf u}\propto({\bf m}_l-{\bf m}_r)$ and $\delta\mu$ (and,
therefore, $\Delta_{\text{in}}$) depends on the polarization $p$ and
relative angle $\theta$ through the combination $p\sin\theta/2$.
In other words, increasing angle with respect to the A alignment at a fixed bias and temperature has the same effect on the thermodynamic properties of the island as decreasing the polarization while keeping the alignment fixed.

The result for the quasiparticle current can than be written similarly to the elastic case (\ref{Iel}) as
\begin{eqnarray}
I_{\text{in}}(\theta,V,T)&=&\left[1-p^{2}\sin^2\theta/2\right] \nonumber \\
&&\times I_0\left(V,T,\Delta_{\text{in}}(V,T,p\sin\theta/2)\right) \, .
\label{Iin}
\end{eqnarray}
The functional form of the gap $\Delta_{\text{in}}(V,T,p)$ has been described in detail in Ref.~\onlinecite{Takahashi:prl99} for the antiparallel alignment and this can be used for the general situation on letting $p\rightarrow p\sin\theta/2$. Comparing Eqs.~(\ref{Iel}) and (\ref{Iin}), one can see that the only difference between the elastic and inelastic regimes in this context is the different superconducting gap $\Delta_{\text{el}}$ and $\Delta_{\text{in}}$, respectively. While in the elastic regime, gap suppression is independent of the relative angle $\theta$, in the inelastic regime, it depends on the angle through the combination $p\sin\theta/2$. Keeping this distinction in mind, we can drop the arguments in Eqs.~(\ref{Iel}) and (\ref{Iin}) and rewrite them as
\begin{equation}
I=[1-p^2\sin^2\theta/2]I_0 \, .
\label{II}
\end{equation}

An important observation to make is that for a given junction conductance, magnetization configuration, voltage bias and temperature, the gap suppression is greater in the elastic than in the inelastic regimes. This follows from Eq.~(\ref{Trf}) for the elastic case, Eq.~(\ref{Trfi}) for the inelastic case, the self-consistency requirement (\ref{sc}) and the fact that $\delta\mu<eV/2$.\cite{Takahashi:prl99}

We can summarize the results in the elastic (\ref{Iel}) and inelastic
regimes (\ref{Iin}) by noting that the gap suppression and the tunnel
conductance of a symmetric \textit{F-S-F} system in the noncollinear alignment
can be found by calculating the corresponding values in the A alignment and
using
\begin{equation}
p_{\text{eff}}=p\sin\theta/2
\label{peff}
\end{equation}
instead of $p$. It is true for both elastic and inelastic
regimes. In particular, Eq.~(\ref{peff}) implies that in the inelastic
regime, the superconducting island undergoes a superconductor--normal
metal phase transition on varying the relative magnetization direction
gradually from the P to A alignment at a fixed voltage.
This occurs when the ferromagnet polarization is large enough so that the
superconducting gap vanishes in the A configuration.

Spin-polarized current driven through a magnetic multilayer system can induce torques on the magnetic layers due to the spin transfer. Many new phenomena related to these current-driven spin torques in magnetic layered systems were both predicted and observed in recent years.\cite{Sloncz:mmm95,Myers:sc00,Sloncz:mmm99} Related to our subject, for example, is torque on the ferromagnet layers in five-layer \textit{N-F-N-F-N} systems studied in Ref.~\onlinecite{Sloncz:mmm95}, where it was shown how the spin torque can be responsible for the mesoscopic precession and switching of the magnetization directions of the ferromagnet layers.

We calculate the torque on the ferromagnets when the spin current passes through the double tunnel barrier \textit{F-S-F} system. The torque exerted on the left ferromagnet is
\begin{equation}
\text{\boldmath$\tau$}=\hat{\bf I}_s-(\hat{\bf I}_s\cdot{\bf m}_l){\bf m}_l \, .
\label{tau}
\end{equation}
First, we find the spin current $\hat{\bf I}_s$ for the \textit{F-S-F}
system (by spin conservation on the node, it is the same in
both junctions):
\begin{equation}
\hat{\bf I}_s=-({\bf m}_r+{\bf m}_l)\frac{\hbar p}{4e}I_0 \, .
\label{Is}
\end{equation}
Here, $I_0$ is given by Eq.~(\ref{I0}),
and the gap $\Delta$ has to be evaluated self-consistently,
as described above for the elastic and inelastic transport regimes. 
The sign convention for the current and the spin current in Eq.~(\ref{Is})
is described in the first paragraph of Section~\ref{Model}.

From Eqs.~(\ref{tau}) and (\ref{Is}) it follows that
\begin{equation}
\text{\boldmath$\tau$}=\hat{\text{\boldmath$\tau$}}\frac{\hbar p \sin\theta}{4e}I_0 \, ,
\label{ntau}
\end{equation}
where the unit vector
$\hat{\text{\boldmath$\tau$}}=({\bf m}_l\cos\theta-{\bf m}_r)/\sin\theta$.

In the elastic voltage-biased system, the gap $\Delta$ and, consequently, $I_0$ (\ref{I0}) do not depend on $\theta$. Therefore, the torque $\tau(\theta)$ (\ref{ntau}) has a simple sinusoidal dependence on the relative angle $\theta$; in particular, $\tau(\theta)$ reaches its maximum when $\theta=\pi/2$. However, in the inelastic transport regime, the gap suppression is larger with larger $p_{\text{eff}}$:\cite{Takahashi:prl99} The gap is most suppressed in the A configuration ($p_{\text{eff}}=p$) and is not suppressed at all (with respect to its equilibrium value $\Delta(T)$) in the P configuration ($p_{\text{eff}}=0$). On the other hand, according to Eq.~(\ref{I0}), higher gap suppression means higher current $I_0$ and, consequently, higher torque $\tau\propto I_0$ (\ref{ntau}). As a result, the maximum in torque $\tau(\theta)$ is shifted to $\theta>\pi/2$ in this regime. In general, for a given junction conductance, magnetization configuration, voltage bias and temperature, the torque will be larger in the elastic than in the inelastic regime as the gap suppression is greater in the elastic regime.

The torque-to-current ratio can be found by using Eqs.~(\ref{ntau}) and (\ref{II}):
\begin{equation}
\tau/I=\frac{\hbar}{4e}\frac{p\sin\theta}{1-p^2\sin^2\theta/2} \, .
\label{nntau}
\end{equation}
In the case of semimetallic contacts, $p=1$, Eq.~(\ref{nntau}) is the same as the result of Slonczewski\cite{Sloncz:mmm95} for \textit{N-F-N-F-N} systems:
\[
\tau/I=\frac{\hbar}{2e}\tan\theta/2 \, ,
\]
even though the angular dependence of the current $I$ is different in our model.

\section{Conclusions and discussions}
\label{Conclusions}

Spin accumulation is crucial for the superconducting gap suppression
in the inelastic regime.\cite{Takahashi:prl99} The mechanism of this
suppression was discussed in detail in
Ref.~\onlinecite{Takahashi:prl99} for the case of antiparallel
alignment and we generalized this to noncollinear magnetization
configuration.

In contrast to this, in the elastic regime, spin accumulation does not
affect the superconducting gap. This follows from Eq.~(\ref{Trf}):
even though the distribution of the quasiparticles is anisotropic in
the spin space for finite ferromagnet polarization $p$, the polarization
$p$ does not affect the total distribution of the quasiparticles in the
momentum space. This is a direct consequence of the charge
conservation on the superconducting island at every energy level. In
this transport regime, the current is proportional to the current in a
similar \textit{N-S-N} system and the relative current change upon changing the
magnetization directions is identical to the situation of a normal
metal island.

We derived a very simple result for the spin torque in \textit{F-S-F} systems (\ref{ntau}) which can be viewed as a generalization of the result for \textit{F-N-F} systems. In the elastic case, the relative torque change upon changing the angle $\theta$ is identical to the situation of \textit{F-N-F} systems. In the inelastic case, the angular dependence of the torque becomes more complicated as the gap suppression is angular-dependent. The maximum torque sets in for $\theta>\pi/2$ due to the suppression of the gap that increases the spin current.

Let us briefly discuss generalizations to go beyond the tunnel regime.
It has been shown\cite{Brataas:prl00} that
Eq.~(\ref{iN}) is the general expression for the current through an
arbitrary \textit{N-F} junction when the spin-flip
relaxation processes can be disregarded.
We believe that Eq.~(\ref{iS}) also holds generally for the
quasiparticle current $\hat{\imath}_{ \text{qp}}$ in
\textit{S-F} systems with arbitrary junctions if
we redefine $N_S(\epsilon)$ to be an energy-dependent normalization factor,
which equals normalized BCS density of states for the tunnel contacts but
has to be found separately in the case of other contacts.
For higher transparency junctions, Andreev reflection also
participates in the transport properties. The Andreev current is
spinless since a Cooper pair consists of an electron and a hole with
opposite spins,
\[
\hat{\imath}_{A}(\epsilon)e=\hat{1}g_A(\epsilon/\Delta)\left(\mbox{Tr}\left[\hat{f}_F(\epsilon)+\hat{f}_F(-\epsilon)\right]-2\right) \, ,
\]
where $g_A$ is an energy-dependent conductance.
The total current is given by
$\hat{\imath}=\hat{\imath}_{\text{qp}}+\hat{\imath} _{A}$.

In conclusion, we have discussed the relevance of the elastic and inelastic transport regimes and showed that at least some experiments on \textit{F-S-F} systems with relatively clean s-wave superconducting island can be in the elastic regime.

We generalized the quasiparticle current for \textit{N-F} tunnel junctions
(\ref{iN}) to describe \textit{S-F} tunnel junctions
by defining an effective difference in the distribution function across
the junction (\ref{fS}). We then used it to find the thermodynamic gap
suppression, electric current-voltage characteristics and
mechanical spin-torque properties of symmetric double barrier \textit{F-S-F}
systems biased by an applied voltage source.  We considered both
the elastic and the inelastic transport regimes. The inelastic regime has been
described in detail elsewhere\cite{Takahashi:prl99} in the case of
collinear magnetization directions of the ferromagnetic reservoirs and
here we generalized it to the case of arbitrary alignment of the
ferromagnets. Angular dependence of the superconducting gap
suppression and current-voltage characteristics can be succinctly
described using effective ferromagnet polarization (\ref{peff}). In
the inelastic transport regime, Eq.~(\ref{peff}) can be used to determine the angle of the phase transition from the normal to superconducting state when the alignment is gradually varied from the antiparallel to parallel at a fixed voltage bias. This occurs when the polarization is large enough so that the island superconductivity is completely suppressed in the antiparallel configuration.

\acknowledgments

We are grateful to G.\ E.\ W.\ Bauer, W.\ Belzig, B.\ I.\ Halperin, D. Huertas Hernando and Yu.\ V.\ Nazarov for stimulating discussions. This work was supported in part by the Norwegian Research Council, NSF Grant No. DMR 99-81283, the Schlumberger Foundation, and the NEDO International Joint Research Grant Program ``Nano-magnetoelectronics''.


\begin{thebibliography}{28}
\expandafter\ifx\csname natexlab\endcsname\relax\def\natexlab#1{#1}\fi
\expandafter\ifx\csname bibnamefont\endcsname\relax
  \def\bibnamefont#1{#1}\fi
\expandafter\ifx\csname bibfnamefont\endcsname\relax
  \def\bibfnamefont#1{#1}\fi
\expandafter\ifx\csname citenamefont\endcsname\relax
  \def\citenamefont#1{#1}\fi
\expandafter\ifx\csname url\endcsname\relax
  \def\url#1{\texttt{#1}}\fi
\expandafter\ifx\csname urlprefix\endcsname\relax\def\urlprefix{URL }\fi
\providecommand{\bibinfo}[2]{#2}
\providecommand{\eprint}[2][]{\url{#2}}

\bibitem[{\citenamefont{Barnas and Fert}(1998)}]{Barnas:epl98}
\bibinfo{author}{\bibfnamefont{J.}~\bibnamefont{Barnas}} \bibnamefont{and}
  \bibinfo{author}{\bibfnamefont{A.}~\bibnamefont{Fert}},
  \bibinfo{journal}{Europhys. Lett.} \textbf{\bibinfo{volume}{44}},
  \bibinfo{pages}{85} (\bibinfo{year}{1998});
  \bibinfo{author}{\bibfnamefont{A.}~\bibnamefont{Brataas}},
  \bibinfo{author}{\bibfnamefont{Yu.~V.}~\bibnamefont{Nazarov}},
  \bibinfo{author}{\bibfnamefont{J.}~\bibnamefont{Inoue}}, \bibnamefont{and}
  \bibinfo{author}{\bibfnamefont{G.~E.~W.} \bibnamefont{Bauer}},
  \bibinfo{journal}{Phys. Rev. B} \textbf{\bibinfo{volume}{59}},
  \bibinfo{pages}{93} (\bibinfo{year}{1999});
  \bibinfo{author}{\bibfnamefont{H.}~\bibnamefont{Imamura}},
  \bibinfo{author}{\bibfnamefont{S.}~\bibnamefont{Takahashi}},
  \bibnamefont{and} \bibinfo{author}{\bibfnamefont{S.}~\bibnamefont{Maekawa}},
  \bibinfo{journal}{Phys. Rev. B} \textbf{\bibinfo{volume}{59}},
  \bibinfo{pages}{6017} (\bibinfo{year}{1999});
  \bibinfo{author}{\bibfnamefont{A.}~\bibnamefont{Brataas}} \bibnamefont{and}
  \bibinfo{author}{\bibfnamefont{X.}~\bibnamefont{Wang}},
  \bibinfo{journal}{Phys. Rev. B} \textbf{\bibinfo{volume}{64}},
  \bibinfo{pages}{104434}  (\bibinfo{year}{2001});
  \bibinfo{author}{\bibfnamefont{A.}~\bibnamefont{Brataas}},
  \bibinfo{author}{\bibfnamefont{M.}~\bibnamefont{Hirano}},
  \bibinfo{author}{\bibfnamefont{J.}~\bibnamefont{Inoue}},
  \bibinfo{author}{\bibfnamefont{Yu.~V.}~\bibnamefont{Nazarov}}, \bibnamefont{and}
  \bibinfo{author}{\bibfnamefont{G.~E.~W.} \bibnamefont{Bauer}},
  \bibinfo{journal}{Jpn. J. appl. Phys.} \textbf{\bibinfo{volume}{40}},
  \bibinfo{pages}{2329} (\bibinfo{year}{2001});
  \bibinfo{author}{\bibfnamefont{G.}~\bibnamefont{Usaj}} \bibnamefont{and}
  \bibinfo{author}{\bibfnamefont{H.~U.}~\bibnamefont{Baranger}},
  \bibinfo{journal}{Phys. Rev. B} \textbf{\bibinfo{volume}{63}},
  \bibinfo{pages}{184418} (\bibinfo{year}{2001}).

\bibitem[{\citenamefont{Brataas
  et~al.}(1999)\citenamefont{Brataas, Nazarov, Inoue, and
  Bauer}}]{Brataas:epjb99}
\bibinfo{author}{\bibfnamefont{A.}~\bibnamefont{Brataas}},
  \bibinfo{author}{\bibfnamefont{Yu.~V.}~\bibnamefont{Nazarov}},
  \bibinfo{author}{\bibfnamefont{J.}~\bibnamefont{Inoue}}, \bibnamefont{and}
  \bibinfo{author}{\bibfnamefont{G.~E.~W.} \bibnamefont{Bauer}},
  \bibinfo{journal}{Eur. Phys. J. B} \textbf{\bibinfo{volume}{9}},
  \bibinfo{pages}{421} (\bibinfo{year}{1999}).

\bibitem[{\citenamefont{Tserkovnyak and Brataas}(2001)}]{Tserkovnyak:prb01}
\bibinfo{author}{\bibfnamefont{Y.}~\bibnamefont{Tserkovnyak}} \bibnamefont{and}
  \bibinfo{author}{\bibfnamefont{A.}~\bibnamefont{Brataas}},
  \bibinfo{journal}{Phys. Rev. B} \textbf{\bibinfo{volume}{64}},
  \bibinfo{pages}{214402} (\bibinfo{year}{2001}).

\bibitem[{\citenamefont{Takahashi et~al.}(1999)\citenamefont{Takahashi,
  Imamura, and Maekawa}}]{Takahashi:prl99}
\bibinfo{author}{\bibfnamefont{S.}~\bibnamefont{Takahashi}},
  \bibinfo{author}{\bibfnamefont{H.}~\bibnamefont{Imamura}}, \bibnamefont{and}
  \bibinfo{author}{\bibfnamefont{S.}~\bibnamefont{Maekawa}},
  \bibinfo{journal}{Phys. Rev. Lett.} \textbf{\bibinfo{volume}{82}},
  \bibinfo{pages}{3911} (\bibinfo{year}{1999}).

\bibitem[{\citenamefont{Slonczewski}(1995)}]{Sloncz:mmm95}
\bibinfo{author}{\bibfnamefont{J.~C.} \bibnamefont{Slonczewski}},
  \bibinfo{journal}{J. Magn. Magn. Mater.} \textbf{\bibinfo{volume}{159}},
  \bibinfo{pages}{L1} (\bibinfo{year}{1995}).

\bibitem[{\citenamefont{Myers et~al.}(2000)\citenamefont{Myers, Ralph, Katine,
  Louie, and Buhrman}}]{Myers:sc00}
\bibinfo{author}{\bibfnamefont{E.~B.} \bibnamefont{Myers}},
  \bibinfo{author}{\bibfnamefont{D.~C.} \bibnamefont{Ralph}},
  \bibinfo{author}{\bibfnamefont{J.~A.} \bibnamefont{Katine}},
  \bibinfo{author}{\bibfnamefont{R.~N.} \bibnamefont{Louie}}, \bibnamefont{and}
  \bibinfo{author}{\bibfnamefont{R.~A.} \bibnamefont{Buhrman}},
  \bibinfo{journal}{Science} \textbf{\bibinfo{volume}{285}},
  \bibinfo{pages}{867} (\bibinfo{year}{2000}).

\bibitem[{\citenamefont{Kaplan et~al.}(1976)\citenamefont{Kaplan, Chi,
  Langenberg, Chang, Jafarey, , and Scalapino}}]{Kaplan:prb76}
\bibinfo{author}{\bibfnamefont{S.~B.} \bibnamefont{Kaplan}},
  \bibinfo{author}{\bibfnamefont{C.~C.} \bibnamefont{Chi}},
  \bibinfo{author}{\bibfnamefont{D.~N.} \bibnamefont{Langenberg}},
  \bibinfo{author}{\bibfnamefont{J.~J.} \bibnamefont{Chang}},
  \bibinfo{author}{\bibfnamefont{S.}~\bibnamefont{Jafarey}}, ,
  \bibnamefont{and} \bibinfo{author}{\bibfnamefont{D.~J.}
  \bibnamefont{Scalapino}}, \bibinfo{journal}{Phys. Rev. B}
  \textbf{\bibinfo{volume}{14}}, \bibinfo{pages}{4854} (\bibinfo{year}{1976}).

\bibitem[{\citenamefont{Al'tshuler and Aronov}(1979)}]{Altshuler:jetp79}
\bibinfo{author}{\bibfnamefont{B.~L.} \bibnamefont{Al'tshuler}}
  \bibnamefont{and} \bibinfo{author}{\bibfnamefont{A.~G.}
  \bibnamefont{Aronov}}, \bibinfo{journal}{Pis'ma Zh. Eksp. Teor. Fiz.}
  \textbf{\bibinfo{volume}{30}}, \bibinfo{pages}{514} (\bibinfo{year}{1979}),
  \bibinfo{note}{[JETP LEtt. {\bf 30}, 482 (1979)]}.

\bibitem[{\citenamefont{Meservey and Tedrow}(1994)}]{Meservey:pr94}
\bibinfo{author}{\bibfnamefont{R.}~\bibnamefont{Meservey}} \bibnamefont{and}
  \bibinfo{author}{\bibfnamefont{P.~M.} \bibnamefont{Tedrow}},
  \bibinfo{journal}{Phys. Rep.} \textbf{\bibinfo{volume}{238}},
  \bibinfo{pages}{173} (\bibinfo{year}{1994}).

\bibitem[{\citenamefont{Johnson and Silsbee}(1988)}]{Johnson:prl88}
\bibinfo{author}{\bibfnamefont{M.}~\bibnamefont{Johnson}} \bibnamefont{and}
  \bibinfo{author}{\bibfnamefont{R.~H.} \bibnamefont{Silsbee}},
  \bibinfo{journal}{Phys. Rev. Lett.} \textbf{\bibinfo{volume}{60}},
  \bibinfo{pages}{377} (\bibinfo{year}{1988}).

\bibitem[{\citenamefont{Yang et~al.}(1994)\citenamefont{Yang, Holody, Lee,
  Henry, Loloee, Schroeder, Pratt, and Bass}}]{Yang:prl94}
\bibinfo{author}{\bibfnamefont{Q.}~\bibnamefont{Yang}},
  \bibinfo{author}{\bibfnamefont{P.}~\bibnamefont{Holody}},
  \bibinfo{author}{\bibfnamefont{S.-F.} \bibnamefont{Lee}},
  \bibinfo{author}{\bibfnamefont{L.~L.} \bibnamefont{Henry}},
  \bibinfo{author}{\bibfnamefont{R.}~\bibnamefont{Loloee}},
  \bibinfo{author}{\bibfnamefont{P.~A.} \bibnamefont{Schroeder}},
  \bibinfo{author}{\bibfnamefont{W.~P.} \bibnamefont{Pratt}}, \bibnamefont{and}
  \bibinfo{author}{\bibfnamefont{J.}~\bibnamefont{Bass}},
  \bibinfo{journal}{Phys. Rev. Lett.} \textbf{\bibinfo{volume}{72}},
  \bibinfo{pages}{3274} (\bibinfo{year}{1994});
  \bibinfo{author}{\bibfnamefont{S.-Y.} \bibnamefont{Hsu}},
  \bibinfo{author}{\bibfnamefont{P.}~\bibnamefont{Holody}},
  \bibinfo{author}{\bibfnamefont{R.}~\bibnamefont{Loloee}},
  \bibinfo{author}{\bibfnamefont{J.~M.} \bibnamefont{Rittner}},
  \bibinfo{author}{\bibfnamefont{W.~P.} \bibnamefont{Pratt}}, \bibnamefont{and}
  \bibinfo{author}{\bibfnamefont{P.~A.} \bibnamefont{Schroeder}},
  \bibinfo{journal}{Phys. Rev. B} \textbf{\bibinfo{volume}{54}},
  \bibinfo{pages}{9027} (\bibinfo{year}{1996}).

\bibitem[{\citenamefont{Abrikosov and Gor'kov}(1962)}]{Abrikosov:zetf62}
\bibinfo{author}{\bibfnamefont{A.}~\bibnamefont{Abrikosov}} \bibnamefont{and}
  \bibinfo{author}{\bibfnamefont{L.~P.} \bibnamefont{Gor'kov}},
  \bibinfo{journal}{Zh. Eksp. Teor. Fiz.} \textbf{\bibinfo{volume}{42}},
  \bibinfo{pages}{1088} (\bibinfo{year}{1962}), \bibinfo{note}{[Sov. Phys. JETP
  \textbf{15}, 1962 (752-757)]}.

\bibitem[{\citenamefont{Meservey and Tedrow}(1978)}]{Meservey:prl78}
\bibinfo{author}{\bibfnamefont{R.}~\bibnamefont{Meservey}} \bibnamefont{and}
  \bibinfo{author}{\bibfnamefont{P.~M.} \bibnamefont{Tedrow}},
  \bibinfo{journal}{Phys. Rev. Lett.} \textbf{\bibinfo{volume}{41}},
  \bibinfo{pages}{805} (\bibinfo{year}{1978}).

\bibitem[{\citenamefont{Vas'ko et~al.}(1997)\citenamefont{Vas'ko, Larkin,
  Kraus, Nikolaev, Grupp, Nordman, and Goldman}}]{Vasko:prl97}
\bibinfo{author}{\bibfnamefont{V.~A.} \bibnamefont{Vas'ko}},
  \bibinfo{author}{\bibfnamefont{V.~A.} \bibnamefont{Larkin}},
  \bibinfo{author}{\bibfnamefont{P.~A.} \bibnamefont{Kraus}},
  \bibinfo{author}{\bibfnamefont{K.~R.} \bibnamefont{Nikolaev}},
  \bibinfo{author}{\bibfnamefont{D.~E.} \bibnamefont{Grupp}},
  \bibinfo{author}{\bibfnamefont{C.~A.} \bibnamefont{Nordman}},
  \bibnamefont{and} \bibinfo{author}{\bibfnamefont{A.~M.}
  \bibnamefont{Goldman}}, \bibinfo{journal}{Phys. Rev. Lett.}
  \textbf{\bibinfo{volume}{78}}, \bibinfo{pages}{1134} (\bibinfo{year}{1997});
  \bibinfo{author}{\bibfnamefont{Z.~W.} \bibnamefont{Dong}},
  \bibinfo{author}{\bibfnamefont{R.}~\bibnamefont{Ramesh}},
  \bibinfo{author}{\bibfnamefont{T.}~\bibnamefont{Venkatesan}},
  \bibinfo{author}{\bibfnamefont{M.}~\bibnamefont{Johnson}},
  \bibinfo{author}{\bibfnamefont{Z.~Y.} \bibnamefont{Chen}},
  \bibinfo{author}{\bibfnamefont{S.~P.} \bibnamefont{Pai}},
  \bibinfo{author}{\bibfnamefont{V.}~\bibnamefont{Talyansky}},
  \bibinfo{author}{\bibfnamefont{R.~P.} \bibnamefont{Sharma}},
  \bibinfo{author}{\bibfnamefont{R.}~\bibnamefont{Shreekala}},
  \bibinfo{author}{\bibfnamefont{C.~J.} \bibnamefont{Lobb}},
  \bibnamefont{et~al.}, \bibinfo{journal}{Appl. Phys. Lett.}
  \textbf{\bibinfo{volume}{71}}, \bibinfo{pages}{1718} (\bibinfo{year}{1997});
  \bibinfo{author}{\bibfnamefont{N.-C.} \bibnamefont{Yeh}},
  \bibinfo{author}{\bibfnamefont{R.~P.} \bibnamefont{Vasquez}},
  \bibinfo{author}{\bibfnamefont{C.~C.} \bibnamefont{Fu}},
  \bibinfo{author}{\bibfnamefont{A.~V.} \bibnamefont{Samoilov}},
  \bibinfo{author}{\bibfnamefont{Y.}~\bibnamefont{Li}}, \bibnamefont{and}
  \bibinfo{author}{\bibfnamefont{K.}~\bibnamefont{Vakili}},
  \bibinfo{journal}{Phys. Rev. B} \textbf{\bibinfo{volume}{60}},
  \bibinfo{pages}{10522} (\bibinfo{year}{1999}).

\bibitem[{\citenamefont{Yamashita et~al.}(2001)\citenamefont{Yamashita,
  Takahashi, Imamura, and Maekawa}}]{Yamashita:cm01}
\bibinfo{author}{\bibfnamefont{T.}~\bibnamefont{Yamashita}},
  \bibinfo{author}{\bibfnamefont{S.}~\bibnamefont{Takahashi}},
  \bibinfo{author}{\bibfnamefont{H.}~\bibnamefont{Imamura}}, \bibnamefont{and}
  \bibinfo{author}{\bibfnamefont{S.}~\bibnamefont{Maekawa}}
  (\bibinfo{year}{2001}), \bibinfo{note}{cond-mat/0108303}.

\bibitem[{\citenamefont{Brataas et~al.}(2000)\citenamefont{Brataas, Nazarov,
  and Bauer}}]{Brataas:prl00}
\bibinfo{author}{\bibfnamefont{A.}~\bibnamefont{Brataas}},
  \bibinfo{author}{\bibfnamefont{Yu.~V.} \bibnamefont{Nazarov}},
  \bibnamefont{and} \bibinfo{author}{\bibfnamefont{G.~E.~W.}
  \bibnamefont{Bauer}}, \bibinfo{journal}{Phys. Rev. Lett.}
  \textbf{\bibinfo{volume}{84}}, \bibinfo{pages}{2481} (\bibinfo{year}{2000});
  \bibinfo{journal}{Eur. Phys. J. B}
  \textbf{\bibinfo{volume}{22}}, \bibinfo{pages}{99} (\bibinfo{year}{2001}).

\bibitem[{\citenamefont{Heslinga and Klapwijk}(1993)}]{Heslinga:prb93}
\bibinfo{author}{\bibfnamefont{D.~R.} \bibnamefont{Heslinga}} \bibnamefont{and}
  \bibinfo{author}{\bibfnamefont{T.~M.} \bibnamefont{Klapwijk}},
  \bibinfo{journal}{Phys. Rev. B} \textbf{\bibinfo{volume}{47}},
  \bibinfo{pages}{5157} (\bibinfo{year}{1993}).

\bibitem[{\citenamefont{Slonczewski}(1999)}]{Sloncz:mmm99}
\bibinfo{author}{\bibfnamefont{J.~C.} \bibnamefont{Slonczewski}},
  \bibinfo{journal}{J. Magn. Magn. Mater.} \textbf{\bibinfo{volume}{195}},
  \bibinfo{pages}{L261} (\bibinfo{year}{1999});
  \bibinfo{author}{\bibfnamefont{X.}~\bibnamefont{Waintal}},
  \bibinfo{author}{\bibfnamefont{E.~B.}~\bibnamefont{Myers}},
  \bibinfo{author}{\bibfnamefont{P.~W.}~\bibnamefont{Brouwer}}, \bibnamefont{and}
  \bibinfo{author}{\bibfnamefont{D.~C.}~\bibnamefont{Ralph}},
  \bibinfo{journal}{Phys. Rev. B} \textbf{\bibinfo{volume}{62}},
  \bibinfo{pages}{12317} (\bibinfo{year}{2000}).

\end{thebibliography}
\end{document}